\documentclass{aa}
\usepackage[varg]{txfonts}
\usepackage{natbib}
\usepackage{url}
\usepackage{hyperref}
\usepackage{xcolor}
\usepackage{ulem}
\usepackage{orcidlink}
\usepackage{xspace}

\usepackage{upgreek}
\usepackage{graphicx}   
\usepackage{multirow}
\usepackage[switch]{linenoaa}
\usepackage{subcaption} 

\newcommand{\src}{{1A~1118$-$61}\xspace}
\newcommand{\nustar}{\textit{NuSTAR}\xspace}

\newcommand{\hxmt}{\textit{Insight-HXMT}\xspace}

\newcommand{\srg}{\textit{SRG}\xspace}

\def\flux{erg\,s$^{-1}$\,cm$^{-2}$\xspace}
\def\lum{erg\,s$^{-1}$\xspace}

\begin{document}

\title{Detection and luminosity-dependent evolution of the high-energy hump in the Be/X-ray pulsar \src}

\titlerunning{High-energy hump in \src}

\authorrunning{Salganik, A., et al.}
\author{Alexander~Salganik\inst{\ref{in:UTU}}\orcidlink{0000-0003-2609-8838}, 
Sergey~S.~Tsygankov\inst{\ref{in:UTU},\ref{in:IHEP}}\orcidlink{0000-0002-9679-0793} 
\and    Sergey~V.~Molkov\inst{\ref{in:IKI}}\orcidlink{0000-0002-5983-5788}
\and Hua Xiao \inst{\ref{in:UTU}} \orcidlink{0009-0004-1288-4912}
\and QingChang Zhao\inst{\ref{in:IHEP}, \ref{in:UCAS}}\orcidlink{0000-0001-9893-8248}
\and 
\\Long Ji \inst{\ref{in:SYS}}\orcidlink{0000-0001-9599-7285}
\and  Alexander A. Mushtukov\inst{\ref{in:UCL},\ref{in:OXF}}\orcidlink{0000-0003-2306-419X} 
\and Igor Yu. Lapshov \inst{\ref{in:IKI}}\orcidlink{0009-0000-4769-452X}  
\and Alexander~A.~Lutovinov\inst{\ref{in:IKI}}\orcidlink{0000-0002-6255-9972}   
\and Alexey Yu. Tkachenko\inst{\ref{in:IKI}}\orcidlink{0000-0002-7486-1730}
\and Hua Feng  \inst{\ref{in:IHEP}}\orcidlink{0000-0001-7584-6236}
\and Shuang-Nan Zhang \inst{\ref{in:IHEP}}\orcidlink{0000-0001-5586-1017}
\and Xiao-Bo Li\inst{\ref{in:IHEP}}\orcidlink{0000-0003-4585-589X}
\and Shu Zhang\inst{\ref{in:IHEP}} 
\and Juri Poutanen\inst{\ref{in:UTU}}\orcidlink{0000-0002-0983-0049}}
\institute{
Department of Physics and Astronomy, 20014 University of Turku,  Finland 
\label{in:UTU} \\ \email{alsalganik@gmail.com}
\and
State Key Laboratory of Particle Astrophysics, Institute of High Energy Physics, Chinese Academy of Sciences, Beijing 100049, China \label{in:IHEP}
\and Space Research Institute, Russian Academy of Sciences, Profsoyuznaya 84/32, 117997 Moscow, Russia 
\label{in:IKI}
\and 
University of Chinese Academy of Sciences, Chinese Academy of Sciences, Beijing 100049, China
\label{in:UCAS}
\and 
School of Physics and Astronomy, Sun Yat-Sen University, Zhuhai 519082, P.R. China
\label{in:SYS}
\and
Mullard Space Science Laboratory, University College London, Holmbury St. Mary, Surrey RH5 6NT, UK
\label{in:UCL}
\and
Astrophysics, Department of Physics, University of Oxford, Denys Wilkinson Building, Keble Road, Oxford OX1 3RH, UK
\label{in:OXF}
}
\date{Received XXX}

\abstract{
\textit{Context.}
Accreting X-ray pulsars exhibit strong luminosity-dependent changes in their broad-band spectra. At high luminosities, their spectra are usually described by a power-law continuum with a high-energy cutoff, whereas low-luminosity observations have revealed a two-hump spectral morphology.

\textit{Aims.}
We aim to trace the luminosity-dependent spectral evolution of the Be/X-ray
pulsar \src\ and to constrain the luminosity range over which the high-energy
hump becomes clearly distinguishable.

\textit{Methods.}
We use dense \srg/ART-XC and \hxmt\ monitoring, together with three broad-band \nustar\ observations of \src\ obtained during its 2026 outburst, to trace the luminosity-dependent evolution of the spectral shape. The ART-XC data follow the decay from a peak luminosity of $\simeq7\times10^{37}$~\lum\ to a low-luminosity plateau at $\simeq(3$--$8)\times10^{35}$~\lum\ in the 4--35~keV band, while the \nustar\ observations provide broad-band spectra during the bright phase, the decline, and the plateau. We describe the continuum with a phenomenological two-component Comptonization model.

\textit{Results.}
As the source faded, the broad-band continuum developed a distinct high-energy hump, giving rise to a two-hump morphology with broad maxima near $\sim$10~keV and $\sim$30--40~keV. The ART-XC monitoring constrains the transition to this morphology to $L_{4-35}\simeq(0.8$--$1.8)\times10^{36}$~\lum. We also find a break in the luminosity dependence of the flux ratio between the two continuum humps around $L_{4-35}\sim10^{37}$~\lum. A cyclotron line at $\simeq55$~keV is detected in the high-energy hump, with no significant luminosity dependence of its centroid energy. We discuss this behavior in the context of resonant interactions in the magnetized accretion flow.
}


\maketitle

\section{Introduction}

The X-ray spectra of accreting X-ray pulsars (XRPs) change strongly with the mass accretion rate, making these systems useful probes of physical processes in extreme environments. At high luminosities, their broad-band X-ray spectra are usually described by a hard continuum with a high-energy cutoff \citep{Coburn2002, Filippova2005}. 
However, recent observations of some systems at low luminosities, reaching $L_{\rm X}\lesssim10^{35}$~\lum, have shown that their spectra can deviate from this canonical shape and instead exhibit two broad humps: one at a few keV and another at several tens of keV \citep{Tsygankov2019b}.

This behavior has been interpreted as evidence that, in the low-accretion-rate regime, the accretion flow is stopped in the neutron star (NS) atmosphere, where collisions excite electrons to higher Landau levels. Their radiative de-excitation produces cyclotron photons, which are then Comptonized and shape the emergent spectrum \citep{SokolovaLapa2021, Mushtukov2021}. Sources that show such luminosity-dependent spectral changes, especially those with independently measured cyclotron features, therefore provide a useful way to connect the continuum morphology with the magnetic-field-controlled radiative processes.

Constraining the formation of the high-energy hump and its dependence on luminosity requires sources that can be followed over a wide luminosity range. Be/X-ray binaries are particularly well suited for this purpose, since they are commonly observed as transients and their outbursts allow the same source to be traced over several orders of magnitude in luminosity \citep{Reig2011}. 
In systems with known cyclotron lines, the line energy provides an independent estimate of the magnetic-field strength in the emission region \citep{Staubert2019}. This makes it possible to compare the energy of the high-energy hump with the cyclotron energy and to test whether the evolving continuum is related to cyclotron processes.

The transient X-ray source \src was discovered by \textit{Ariel-5} during its 1974 outburst, which lasted about one month \citep{Eyles1975}.
It was identified as an XRP with a spin period of $405.3\pm0.6$~s \citep{Ives1975, Fabian1975}.  
The companion is an O9.5\,IV--Ve Be star \citep{Chevalier1975, JanotPacheco1981}, which identifies the system as a Be/X-ray binary. The distance to the source was estimated to be 3--7~kpc \citep{Coe1985, JanotPacheco1981}. The orbital period, $P_{\rm orb} = 24.0 \pm 0.4$ days, along with other orbital parameters, was reported by \citet{Staubert2011}. 

Two further month-long outbursts were detected in 1992 and 2009, reaching peak fluxes of $\sim$150~mCrab in the 20--100~keV band and $\sim$500~mCrab in the 15--50~keV band, respectively \citep{Coe1994, Mangano2009a, Mangano2009b, Doroshenko2010}. 
The broad-band spectrum showed an absorption feature at $\sim$55~keV, interpreted as a cyclotron line corresponding to magnetic field strength of $4.8 \times 10^{12}$~G \citep{Doroshenko2010, Suchy2011, Maitra2012}. 
For the 2009 outburst, \citet{ReigNespoli2013} reported a peak 3--30~keV luminosity of $\approx 3\times10^{37}$~\lum. They also found that \src\ did not enter the supercritical accretion state, in which the accretion flow is decelerated by radiation pressure in an extended accretion column above the NS surface \citep{BaskoSunyaev1976, Mushtukov2015}.

A new outburst of the Be/XRP \src was detected on 2026 January 17 by the \textit{Swift}/BAT telescope \citep{ATel17608, GCN43442, GCN43435}. 
It was subsequently confirmed by \textit{MAXI}/GSC and \srg/ART-XC observations \citep{ATel17613, Molkov2026}. 
Optical spectroscopy obtained with the Southern African Large Telescope (SALT) on 2026 January 21 revealed strong Balmer emission lines, indicating the presence of an active Be circumstellar disc during the current outburst \citep{ATel17620}.
We obtained three \nustar target-of-opportunity observations during the 2026 outburst, sampling the bright, intermediate, and low-luminosity stages. Preliminary results from the bright-state observation were reported in \citet{ATel17626}.

In this paper, we present an analysis of the 2026 outburst of \src\ using dense \srg/ART-XC and \hxmt\ monitoring, together with three broad-band \nustar\ observations.
The observations and data reduction are described in Sect.~\ref{sec:observations}. 
In Sect.~\ref{sec:results}, we present the luminosity-dependent spectral
evolution. 
The onset of radiative braking at high luminosities, the emergence of the two-hump spectral shape as the source fades, and the behavior of the CRSF energy are discussed in Sect.~\ref{sec:discussion}.
Our main conclusions are summarized in Sect.~\ref{sec:conclusions}.

\section{Observations and data reduction}
\label{sec:observations}

\subsection{\nustar}

\nustar\ comprises two identical, co-aligned X-ray focal plane modules, FPMA and FPMB \citep{Harrison2013}, providing sensitive coverage in the 3--79~keV energy band. 
The source was observed three times during the 2026 outburst: on 2026 January~24 (MJD 61064.1--61064.4; ObsID 91201305002), February~21 (MJD 61092.4--61093.0; ObsID 91201310002), and  March~22 (MJD 61121.8--61123.3; ObsID 91202318002). 
These observations are hereafter referred to as NuObs1, NuObs2, and NuObs3, respectively. 

Source events were extracted from circular regions with radii of $60\arcsec$, $50\arcsec$,  and $50\arcsec$ for NuObs1--3, respectively, centered on the source position. 
Background events were extracted from a nearby source-free circular region with a radius of $120\arcsec$ on the same detector chip. 
Data reduction followed the standard \nustar\ processing guidelines.\footnote{\url{https://heasarc.gsfc.nasa.gov/docs/nustar/analysis/nustar_swguide.pdf}}

The data were processed using \textsc{HEASoft} v6.36 and CALDB version 20260407. 
To mitigate background enhancements during passages through the South Atlantic Anomaly (SAA), we adopted the recommended \texttt{nupipeline} filtering configuration, namely \texttt{saacalc=1}, \texttt{saamode=OPTIMIZED}, and \texttt{tentacle=yes}. 
For the brighter observation (NuObs1), we additionally applied the event filtering expression \texttt{`(STATUS==b0000xxx00xxxx000)\&(SHIELD==0)'} to reduce possible contamination from background events associated with high count rates. 
The remaining effective exposures are 13.2, 27.6, and 58.7~ks for NuObs1--3, respectively.

Spectra and light curves were extracted using the \texttt{nuproducts} task within the \texttt{nustardas} pipeline.
The background-subtracted light curves from FPMA and FPMB were combined using \texttt{lcmath} to improve the signal-to-noise ratio. 

\subsection{\textit{SRG}/ART-XC}

The \srg mission, launched in 2019 from the Baikonur Cosmodrome to the Sun–Earth L2 point, carries two X-ray instruments, one of which is the Mikhail Pavlinsky ART-XC telescope \citep{2021A&A...656A.132S}.
ART-XC consists of seven co-aligned mirror-system--detector modules, each comprising an X-ray mirror system and a CdTe double-sided strip detector \citep{2021A&A...650A..42P}.
The instrument is designed to operate primarily in the 4--30~keV energy range, while retaining sensitivity up to $\sim$120~keV. 
In this work, we use the 4--35~keV band for the ART-XC spectral-flux estimates in order to cover the high-energy hump as fully as possible while retaining adequate instrumental sensitivity.

The dataset was processed using the \textsc{artproducts} v1.0 software package together with the latest available calibration files (CALDB v20230228). 
Source events were selected from a circular extraction region of radius {$2\farcm5$} centered on the target, with additional energy selection applied when required. 
The \srg/ART-XC monitoring of \src during the 2026 outburst consisted of 22 observations obtained between January~19 and  April~1 (ObsIDs: 12610180001--12610180022), see Table~\ref{table:obslist}. 

To estimate the background, we used an annulus region centered on the source with inner and outer radii $4'$ and $7'$, respectively. No other sources were detected within the ART-XC field of view with \src being the only contributor to the measured signal.

\subsection{\hxmt}

The Hard X-ray Modulation Telescope \citep[\hxmt;][]{Zhang2020} is China’s first X-ray astronomy satellite, launched on 2017 June 15.
It carries three scientific payloads: the High Energy instrument \citep[HE; 20--250~keV;][]{Liu2020}, the Medium Energy instrument \citep[ME; 5--30~keV;][]{Cao2020}, and the Low Energy instrument \citep[LE; 1--15~keV;][]{Chen2020}.
These payloads consist of detector units with different fields of view and orientations, with total effective areas of 5000, 952, and 384~cm$^2$ for HE, ME, and LE, respectively.
\hxmt conducted seven pointing observations of \src from 2026 January 21 to 27.

The data reduction was performed by using the \hxmt Data Analysis Software (\textsc{HXMTDAS} v2.07),\footnote{\url{http://hxmtweb.ihep.ac.cn/software.jhtml}} along with the CALDB version 2.08.\footnote{\url{http://hxmtweb.ihep.ac.cn/SoftDoc/847.jhtml}} 
The good time intervals (GTIs) were generated using the \texttt{he/me/legtigen} tools with the following criteria: (1) the elevation angle >10$\degr$; (2) the time to SAA $>300$~s; (3) the cutoff rigidity value $>8$~GV; (4) the Moon angle $>1\degr$; and (5) the Sun angle $>70\degr$. 
The tools \texttt{he/me/lescreen} were used to generate the event data. 
The arrival times of the events were subsequently corrected for the Solar system barycenter by the \texttt{hxbary} tools. 
We extracted light curves and spectra using the tasks \texttt{he/me/lelcgen} and \texttt{he/me/lespecgen}, respectively. 
The background was scaled using the \texttt{he/me/lebkgmap} tasks.

\subsection{Spectral data approximation}
All \nustar, ART-XC, and HXMT spectra were grouped to ensure at least 1, 25, and 25 counts per energy channel, respectively. The \nustar spectra were analyzed using the W-statistic \citep{Wachter1979}, while the ART-XC and HXMT spectra were analyzed using the chi-squared statistic. Spectral fitting was carried out with the \textsc{xspec} package (v12.15.1; \citealt{Arnaud1996}).

For the ART-XC and HXMT spectra, we included systematic uncertainties of 1.5\%
and 1\%, respectively, using the \textsc{xspec} \texttt{systematic} command.  No additional systematic uncertainty was applied to the \nustar spectra.  
For the ART-XC spectra, this systematic uncertainty was added to account for
residual calibration uncertainties. As a result, several ART-XC fits
have formally low reduced $\chi^2$ values (Table~\ref{table:obslist}); these
values are therefore not indicative of overfitting.

Unless stated otherwise, all uncertainties are quoted at the 1$\sigma$ confidence level. All fluxes reported in this work are absorption-corrected (unabsorbed) values unless explicitly noted otherwise. Throughout the paper, we assume a source distance of 5~kpc.
\begin{figure}
\centering
\includegraphics[width=0.9\linewidth]{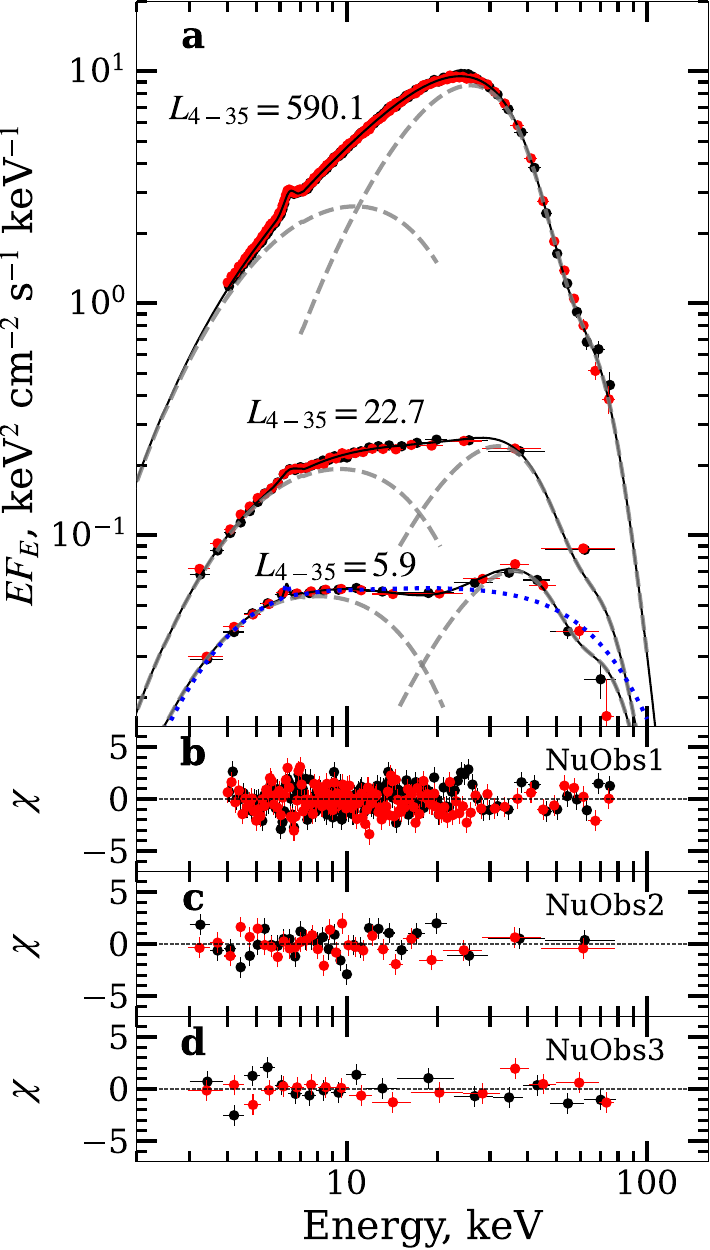}
\caption{
Unfolded \nustar\ spectra of \src. 
The spectra of NuObs1--3 are shown from top to bottom, corresponding to decreasing luminosity. 
The luminosity in the 4--35~keV range $L_{4-35}$ is given in units of $10^{35}$~\lum. 
The FPMA and FPMB data are shown with black and red symbols, respectively. 
The solid line represents the best-fit \texttt{comptt+comptt} continuum model, while the dashed grey lines show the individual low- and high-energy Comptonization components. 
For NuObs3, the blue dotted line shows the best-fit single-\texttt{comptt} continuum model. 
The lower panels show the residuals with respect to the best-fit two-component model for each observation. }

\label{fig:spec}
\end{figure}
\begin{figure}
\centering
\includegraphics[width=0.9\linewidth]{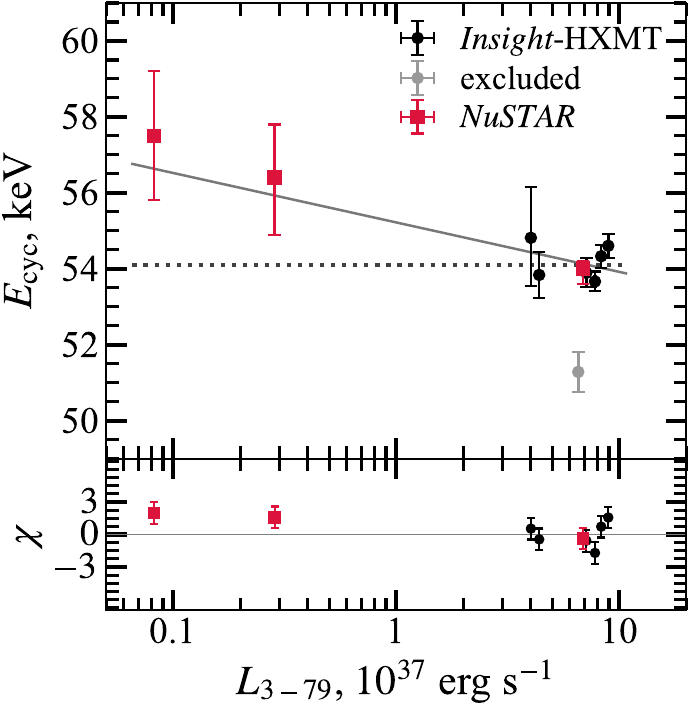}
\caption{Cyclotron-line centroid energy $E_{\rm cyc}$ as a function of the
3--79~keV luminosity $L_{3-79}$. The dotted horizontal line shows the best-fit constant-energy model, while the solid curve shows the best-fit log-linear dependence on $L_{3-79}$. The lower panel shows residuals with respect to the constant model.
}
\label{fig:ecyc_lum}
\end{figure}

\section{Results}
\label{sec:results}

\begin{table*}
\centering
\caption{Best--fit spectral parameters of \src\ obtained with the model 
$\texttt{constant} \times \texttt{tbabs} \times \texttt{mtable\{nuMLIv1.mod\}} \times (\texttt{comptt}_{\rm low}+\texttt{comptt}_{\rm high} + \texttt{gaussian})\times \texttt{gabs}$. }
\label{tab:specs}
\begin{tabular}{lcccccc}
\hline\hline
Parameter & \multicolumn{2}{c}{NuObs1} & \multicolumn{2}{c}{NuObs2} & \multicolumn{2}{c}{NuObs3} \\
 & Low-$E$ & High-$E$ & Low-$E$ & High-$E$ & Low-$E$ & High-$E$ \\
\hline

$C_{\rm FPMA}$ & \multicolumn{2}{c}{1.0 (fixed)} 
               & \multicolumn{2}{c}{1.0 (fixed)}
               & \multicolumn{2}{c}{1.0 (fixed)} \\

$C_{\rm FPMB}$ & \multicolumn{2}{c}{$1.022_{-0.002}^{+0.001}$} 
               & \multicolumn{2}{c}{$1.022\pm0.004$}
               & \multicolumn{2}{c}{$1.024\pm0.006$} \\

$f_\textrm{MLI, FPMA}$ &  \multicolumn{2}{c}{1.0 (fixed)} & \multicolumn{2}{c}{1.0 (fixed)} & \multicolumn{2}{c}{1.0 (fixed)}\\
$f_\textrm{MLI, FPMB}$ &  \multicolumn{2}{c}{$0.62\pm0.03$} &  \multicolumn{2}{c}{$0.63^{+0.04}_{-0.05}$} &  \multicolumn{2}{c}{$0.83\pm0.05$}\\

$N_{\rm H}$ ($10^{22}$ cm$^{-2}$) 
               & \multicolumn{2}{c}{1.1 (fixed)} 
               & \multicolumn{2}{c}{1.1 (fixed)}
               & \multicolumn{2}{c}{1.1 (fixed)} \\

$kT_{0}$ (keV)  
               & \multicolumn{2}{c}{$1.25^{+0.03}_{-0.02}$} 
               & \multicolumn{2}{c}{$1.18\pm0.04$}
               & \multicolumn{2}{c}{$0.98\pm0.03$}\\

$kT_{\rm e}$ (keV) 
               & $3.6\pm0.2$ & $6.7\pm0.1$ 
               & $3.5\pm0.2$ & $8.5\pm0.3$
               & $3.2\pm0.1$ & $10.0\pm0.4$ \\

$\tau$         
               & $7.0\pm0.3$ & 200 (fixed) 
               & $6.4\pm0.4$ & 200 (fixed)
               & $6.3\pm0.4$ & 200 (fixed) \\

$E_{\rm cyc}$ (keV) 
               & \multicolumn{2}{c}{$54.0^{+0.2}_{-0.4}$} 
               & \multicolumn{2}{c}{$56.4^{+1.4}_{-1.5}$}
               & \multicolumn{2}{c}{$57.5\pm1.7$} \\

$\sigma_{\rm cyc}$ (keV) 
               & \multicolumn{2}{c}{$10.7^{+0.9}_{-0.6}$} 
               & \multicolumn{2}{c}{10.7 (fixed)}
               & \multicolumn{2}{c}{10.7 (fixed)} \\

$\tau_{\rm cyc}$ &\multicolumn{2}{c}{$0.7\pm0.1$} & \multicolumn{2}{c}{$0.5\pm0.1$} & \multicolumn{2}{c}{$0.4\pm0.1$}   
                \\

$E_{\rm Fe}$ (keV) 
               & \multicolumn{2}{c}{$6.43\pm0.01$} 
               & \multicolumn{2}{c}{$6.40\pm0.06$}
               & \multicolumn{2}{c}{$6.31\pm0.04$} \\

$\sigma_{\rm Fe}$ (keV) 
               & \multicolumn{2}{c}{$0.25\pm0.01$} 
               & \multicolumn{2}{c}{$0.4\pm0.1$}
               & \multicolumn{2}{c}{0.1 (fixed)} \\

EW (keV)       
               & \multicolumn{2}{c}{$0.097\pm0.003$} 
               & \multicolumn{2}{c}{$0.07\pm0.01$}
               & \multicolumn{2}{c}{$0.04\pm0.01$} \\

Flux$_{3-79}$ ($10^{-10}$~\flux) 
               & \multicolumn{2}{c}{$230.0\pm0.2$} 
               & \multicolumn{2}{c}{$9.53\pm0.04$}
               & \multicolumn{2}{c}{$2.74\pm0.02$} \\

Luminosity$_{3-79}$ ($10^{35}$~\lum) 
               & \multicolumn{2}{c}{$688.0^{+0.6}_{-0.7}$} 
               & \multicolumn{2}{c}{$28.5\pm0.1$}
               & \multicolumn{2}{c}{$8.20\pm0.05$} \\
Luminosity$_{4-35}$ ($10^{35}$~\lum) 
               & \multicolumn{2}{c}{$590.1^{+0.5}_{-0.4}$} 
               & \multicolumn{2}{c}{$22.7\pm0.6$}
               & \multicolumn{2}{c}{$5.94\pm0.02$} \\

W-statistic/d.o.f.  
               & \multicolumn{2}{c}{3296 / 3272} 
               & \multicolumn{2}{c}{2529 / 2729}
               & \multicolumn{2}{c}{2674 / 2738} \\

\hline
\end{tabular}
\end{table*}

\subsection{Broad-band spectral shape}
\label{sec:spectrum}

We performed spectral analysis of the three \nustar\ observations obtained during the outburst, fitting the FPMA and FPMB data simultaneously in the 3--79~keV energy range using \textsc{xspec} (for the brightest observation NuObs1 we limited the range to 4--79~keV). 
We found noticeable discrepancies between the FPMA and FPMB spectra at low energies ($\lesssim 7$~keV), consistent with recent calibration updates for \nustar\ related to changes in the soft X-ray response of FPMB following the degradation of its multi-layer insulation (MLI) after 2024.\footnote{\url{https://nustarsoc.caltech.edu/NuSTAR_Public/NuSTAROperationSite/mli.php}} 

To account for this effect, we applied the MLI correction using the \texttt{nuMLIv1.mod} table model provided by the \nustar\ Science Operations Center. 
Following the recommended procedure, we first excluded the 3--7~keV energy range for FPMB and determined the cross-normalization constant (\texttt{const} spectral model) between FPMA and FPMB. 
After fixing this constant, we restored the excluded energy range and applied the MLI correction to the FPMB spectrum only, while keeping the FPMA MLI fraction fixed at unity.

As a first step, we attempted to describe the spectrum in the lowest-luminosity state (NuObs3; $L_{\rm 3-79}\simeq8\times10^{35}$~\lum) using the Comptonization model \texttt{comptt} \citep{Titarchuk1994}, which is commonly applied in the analysis of accreting XRPs. 
The photoelectric absorption was modeled with the \texttt{tbabs} component adopting the abundances of \citet{Wilms2000}. 
The interstellar absorption column density was fixed at $N_{\rm H}=1.1\times10^{22}~\mathrm{cm^{-2}}$, corresponding to the Galactic value derived from the HI4PI survey in the direction of the source \citep{HI4PI2016}. 
In addition, we included a Gaussian emission line to account for the iron K$\alpha$ feature near 6.4~keV and a multiplicative absorption component (\texttt{gabs}) to model the previously mentioned cyclotron resonant scattering feature (CRSF) at $\sim$55~keV.

However, this model does not adequately reproduce the observed spectral shape, showing significant systematic deviations above $\sim$20~keV (see Fig.~\ref{fig:spec}). 
The unfolded spectrum instead reveals a pronounced double-peaked structure with two broad maxima located at approximately $\sim$10 and $\sim$30--40~keV. 
This morphology is not adequately reproduced by the single-component continuum tested here and is often observed in  XRPs at luminosities $L_{\rm 3-79}\sim10^{34}$--$10^{36}$~\lum\ \citep[e.g.,][]{Tsygankov2019b, Zalot2026}. 

We therefore modeled the continuum as the sum of two thermal Comptonization components \texttt{comptt+comptt}, following \citet{Tsygankov2019b}. 
In each observation, both components were assumed to share the same seed-photon temperature $kT_{0}$. 
The electron temperature and optical depth of the low-energy component were allowed to vary independently. 
For the high-energy component (i.e., the \texttt{comptt} component with the higher electron temperature), the optical depth was fixed at $\tau = 200$ to stabilize the fit and reduce its degeneracy with the electron temperature. 
This value should therefore be regarded as purely phenomenological.
The resulting model is $\texttt{constant} \times \texttt{tbabs} \times \texttt{mtable\{nuMLIv1.mod\}} \times (\texttt{comptt}_{\rm low}+\texttt{comptt}_{\rm high} + \texttt{gaussian})\times \texttt{gabs}$. 
The model provides a statistically acceptable description of the data, with W-statistic/d.o.f. values close to unity, and successfully reproduces both spectral peaks.  
The best-fit spectral parameters are reported in Table~\ref{tab:specs}, and the corresponding spectra are shown in Fig.~\ref{fig:spec}.

At intermediate luminosity (NuObs2; $L_{\rm 3-79}\simeq3\times10^{36}$~\lum), the spectrum retains a less pronounced double-peaked morphology similar to that observed in the lowest-luminosity state. 
In contrast, during the brightest observation (NuObs1; $L_{\rm 3-79}\simeq7\times10^{37}$~\lum), the spectral shape resembles a more typical cutoff power-law continuum, and although the two components are not clearly separated in NuObs1, we retain the same continuum prescription for consistency with the lower-luminosity observations (see Table~\ref{tab:specs}).

\subsection{Cyclotron-line energy as a function of luminosity}
\label{sec:ecyc_lx}

Luminosity-dependent variations of cyclotron-line energies have been reported in several accreting XRPs, but no universal behavior is observed: positive, negative, and absent correlations are known, depending on the source and luminosity regime \citep[see, e.g.,][]{Staubert2019}.
To investigate whether such a dependence is present in \src\ over the luminosity range covered by the 2026 outburst, we supplemented the \nustar\ measurements with seven \hxmt\ observations obtained during the bright stage, at $L_{3-79}\gtrsim4\times10^{37}$~\lum.
The spectra were fitted in the 4--79~keV range with the same phenomenological continuum model as used for \nustar, but without the \nustar-specific MLI correction. 
The CRSF width was fixed at $\sigma_{\rm cyc} = 10.7$~keV, corresponding to the value measured for NuObs1 (Table~\ref{tab:specs}). 
The seed-photon temperature $kT_{0}$ was fixed to luminosity-dependent values interpolated from the three \nustar\ measurements, following the interpolation procedure described for the ART-XC spectra in Sect.~\ref{sec:artxc_lightcurve}. 
The low-energy component electron temperature $kT_{\rm e,low}$ was left free. 
The resulting fit parameters are reported in Table~\ref{table:obslist}.
The cyclotron energies are shown in Fig.~\ref{fig:ecyc_lum}. 

We fitted the combined \nustar\ and \hxmt\ data with a constant-energy model and with a log-linear dependence on luminosity,
$E_{\rm cyc} = a + b \log_{10}\left(L_{3-79}/(10^{37}\ {\rm erg\ s^{-1}})\right)$.
The constant-energy model gives $E_{\rm cyc}=54.1\pm0.1$~keV and provides an acceptable description of the data ($\chi^2/{\rm d.o.f.}=13/8$).
Allowing the line energy to vary log-linearly with luminosity improves the fit only marginally, giving $\chi^2/{\rm d.o.f.}=9/7$.
The best-fit slope is $b=-1.3\pm0.6$~keV~dex$^{-1}$, formally different from zero at the $\simeq2\sigma$ level.
Although this may indicate a weak decrease of the phase-averaged cyclotron energy with increasing luminosity, the statistical significance is low and the constant-energy model remains acceptable.
We therefore do not claim a significant intrinsic luminosity dependence of the phase-averaged cyclotron energy over the range probed by our observations.
This is consistent with the earlier findings for this source reported by \citet{ReigNespoli2013}.

We excluded the HXMT measurement at $L_{3-79}\simeq6.5\times10^{37}$~\lum\ from the luminosity-dependence fit because its cyclotron energy is an isolated outlier relative to the adjacent \hxmt observations. 
The measured CRSF parameters may be affected by systematic uncertainties. 
For the \hxmt\ measurements, these include residual uncertainties in the HE background modeling and possible drifts in the energy--channel relation. In addition, the CRSF centroid energy can depend on the adopted continuum model \citep[e.g.,][]{Staubert2019}. 

\begin{figure}
\centering
\includegraphics[width=0.9\linewidth]{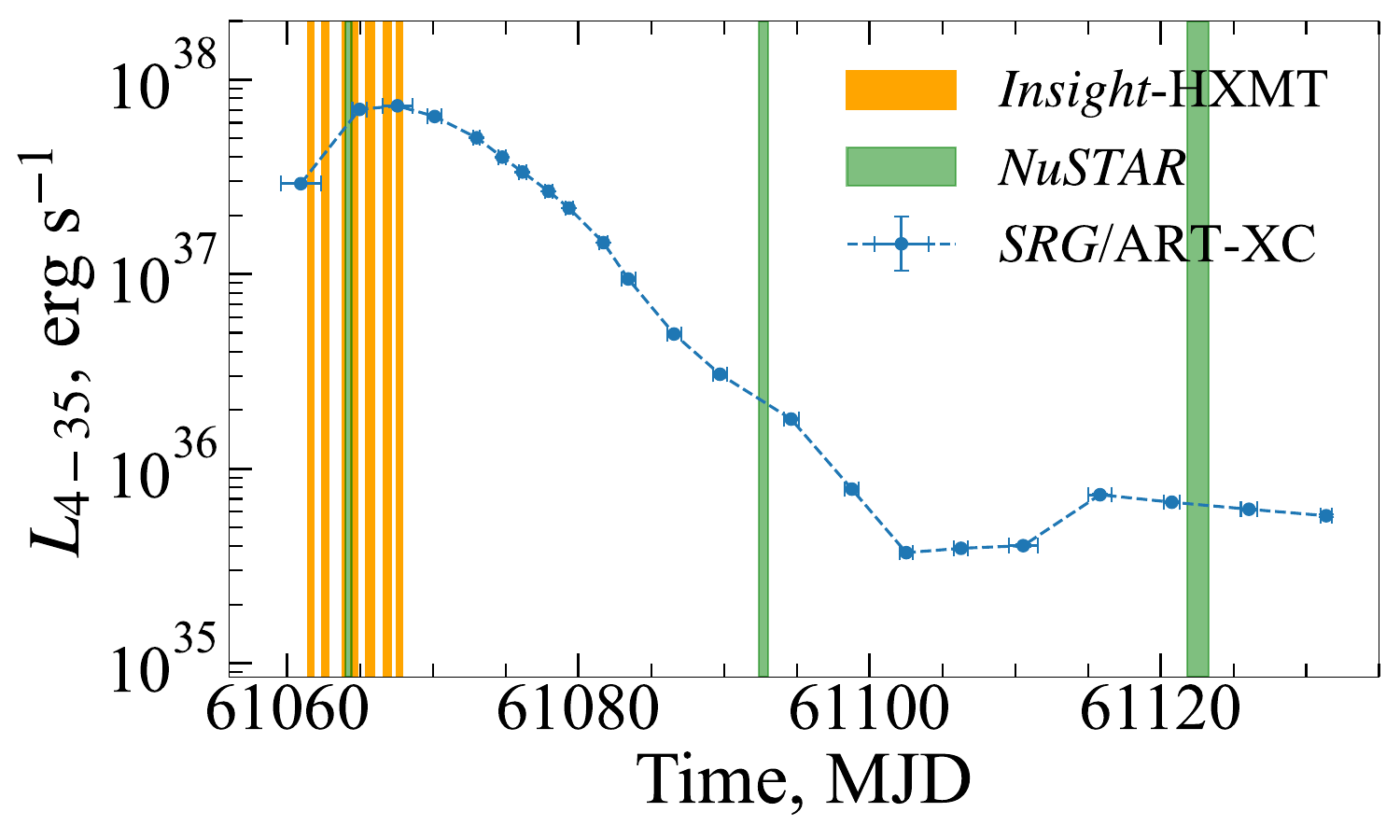}
\caption{\srg/ART-XC light curve of \src in the 4--35~keV energy band. The shaded green regions mark the time intervals of the \nustar observations, while the orange regions indicate the intervals of the \hxmt observations. }
    \label{fig:lightcurve}
\end{figure}

\subsection{The 2026 outburst light curve and spectral evolution}
\label{sec:artxc_lightcurve}

To trace the long-term evolution of the 2026 outburst and to determine the luminosity range over which the high-energy hump becomes clearly distinguishable, we analysed the individual \srg/ART-XC pointed observations. 
For each observation, the source spectrum was fitted independently in the 5--35~keV energy range with a simplified version of the broad-band model described in Sect.~\ref{sec:spectrum}, \texttt{comptt}$_{\rm low}$+\texttt{comptt}$_{\rm high}$+\texttt{gaussian}.
The resulting best-fit models were then used to estimate the unabsorbed
4--35~keV fluxes and to construct the ART-XC light curve.

The \nustar-specific MLI correction table, the cross-normalization constant, and the CRSF component were omitted. 
The CRSF component was excluded because the line lies outside the ART-XC fitting range. 
To stabilize the fits, we fixed the seed-photon temperature $kT_{0}$ and the electron temperature of the low-energy Comptonization component, $kT_{\rm e,low}$, rather than fitting them independently in each ART-XC spectrum.

For each ART-XC spectrum, we first performed a fit with $kT_{0}=1.25$~keV and $kT_{\rm e,low}=3.6$~keV, i.e. the NuObs1 values listed in Table~\ref{tab:specs}. 
This fit was used only to obtain a preliminary estimate of the 4--35~keV luminosity. 
We then used this luminosity to assign fixed values of $kT_{0}$ and $kT_{\rm e,low}$ by linearly interpolating the tabulated NuObs1--NuObs3 values as a function of $\log L_{4-35}$. 
The ART-XC spectra were then refitted with these values fixed to obtain the final luminosities. 
For luminosities outside the range covered by the \nustar observations, the nearest \nustar values were used. 
The iron-line energy and width were fixed at $E_{\rm Fe}=6.4$~keV and
$\sigma_{\rm Fe}=0.1$~keV.

The ART-XC light curve is shown in Fig.~\ref{fig:lightcurve}, and the corresponding spectral parameters and luminosities are listed in
Table~\ref{table:obslist}. It covers the outburst from the early bright phase through the subsequent decay and shows a gradual drop from $\simeq7\times10^{37}$~\lum in the 4--35~keV range. 
After the decline,
the source reached a plateau around MJD 61100 at a level of (3--8$)\times10^{35}$~\lum. 
The three \nustar\ observations sampled different stages of this evolution: NuObs1 was obtained during the bright  phase, NuObs2 after the main decline, close to the low-luminosity state, and NuObs3 during the late low-luminosity stage, when the source had settled at the plateau.

\begin{figure}
\centering
\includegraphics[width=0.9\linewidth]{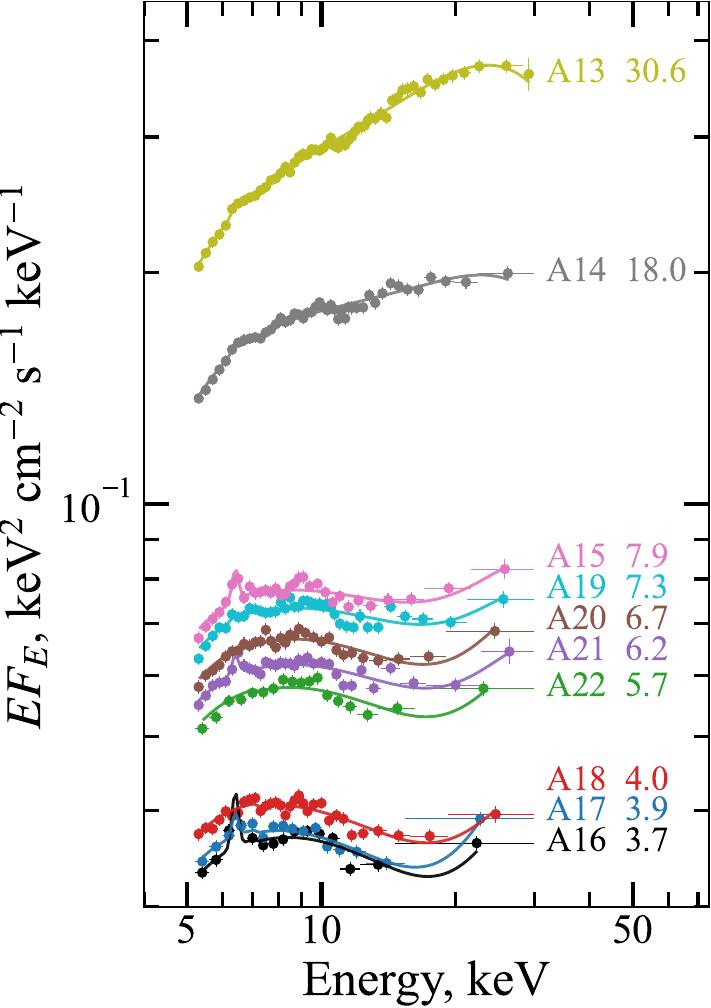}
\caption{Evolution of the unfolded \srg/ART-XC spectra of \src during the low-luminosity stage of the 2026 outburst. Solid curves show the corresponding best-fit spectral models. For clarity, only observations 12610180013--12610180022 (labeled in the figure as A13--A22) are displayed, with their corresponding unabsorbed 4--35~keV luminosities indicated on the right in units of $10^{35}$~\lum. The remaining spectra are not shown for clarity.}
\label{fig:artxc_spectral_evolution}
\end{figure}

As the source fades, the high-energy hump becomes more clearly separated (see Fig.~\ref{fig:artxc_spectral_evolution}). The first ART-XC observation in which the two humps are clearly separated is ObsID 12610180015, labeled as A15 in the figure, at $L_{4-35}\approx8\times10^{35}$~\lum. In the preceding observation, ObsID 12610180014 (A14), at $L_{4-35}\approx1.8\times10^{36}$~\lum, the high-energy hump is not yet clearly resolved as a separate component.

\begin{figure}
\centering
\includegraphics[width=0.9\linewidth]{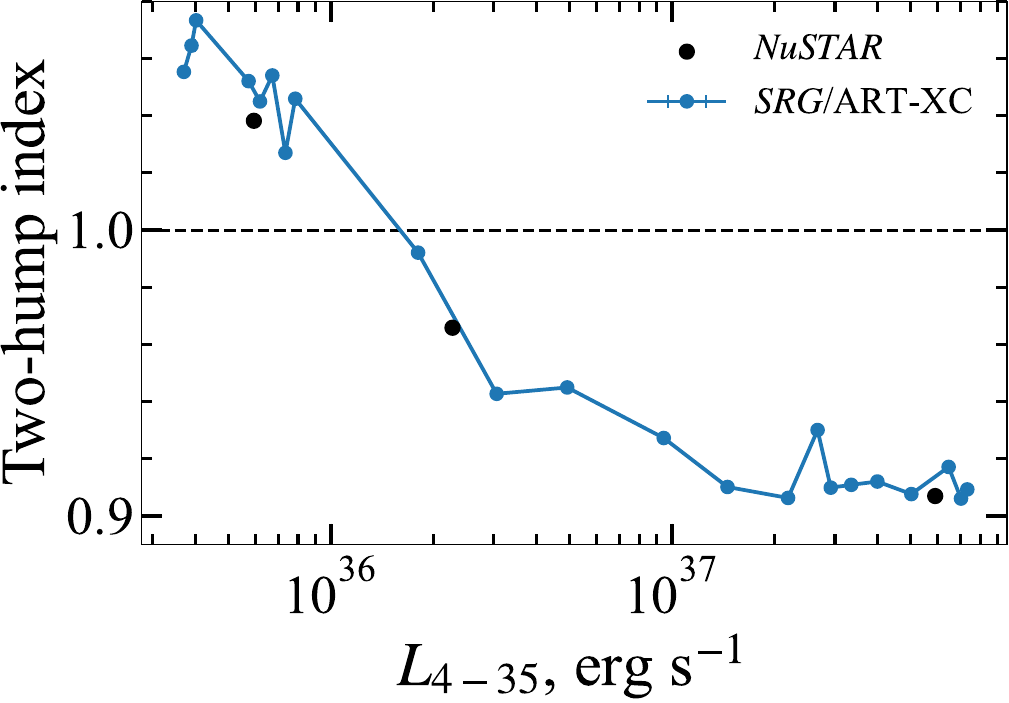}
\caption{
Luminosity dependence of the spectral-shape index $S$  given by Eq.~\eqref{eq:index} and  measured from the interpolated unfolded ART-XC spectra.
Black points show the corresponding values measured from the unfolded \nustar
spectra. The horizontal dashed line marks unity.}
    \label{fig:twohump_shape_index}
\end{figure}

To quantify this, we introduced an empirical spectral-shape index based on the interpolated unfolded ART-XC spectra in the $E F_E$ representation. 
We 
define
\begin{equation} 
\label{eq:index}
S = \frac{0.5\,[Y(10~{\rm keV})+Y(22~{\rm keV})]}{Y(16~{\rm keV})}, 
\end{equation}
where $Y(E)=E F_E$. 
For each observation, $Y(E)$ at 10, 16, and 22~keV was obtained by interpolating the unfolded spectral points. 
The 16~keV energy samples the average intersection between the two \texttt{comptt} components in NuObs2 and NuObs3, while 10 and 22~keV probe the adjacent low- and high-energy parts of the spectrum. 
Values of $S>1$ indicate a depression near 16~keV and therefore a more pronounced
two-hump morphology.

The luminosity dependence of the empirical shape index $S$ is shown in
Fig.~\ref{fig:twohump_shape_index}. 
The index increases as the source fades and first exceeds unity at $L_{4-35}\simeq8\times10^{35}$~\lum, consistent with the first ART-XC observation in which the two-hump morphology is clearly resolved.
Together with the preceding observation, this constrains the
luminosity range over which the high-energy hump becomes clearly resolved to
$L_{4-35}\simeq(0.8$--$1.8)\times10^{36}$~\lum.

\begin{figure}
\centering
\includegraphics[width=0.9\linewidth]{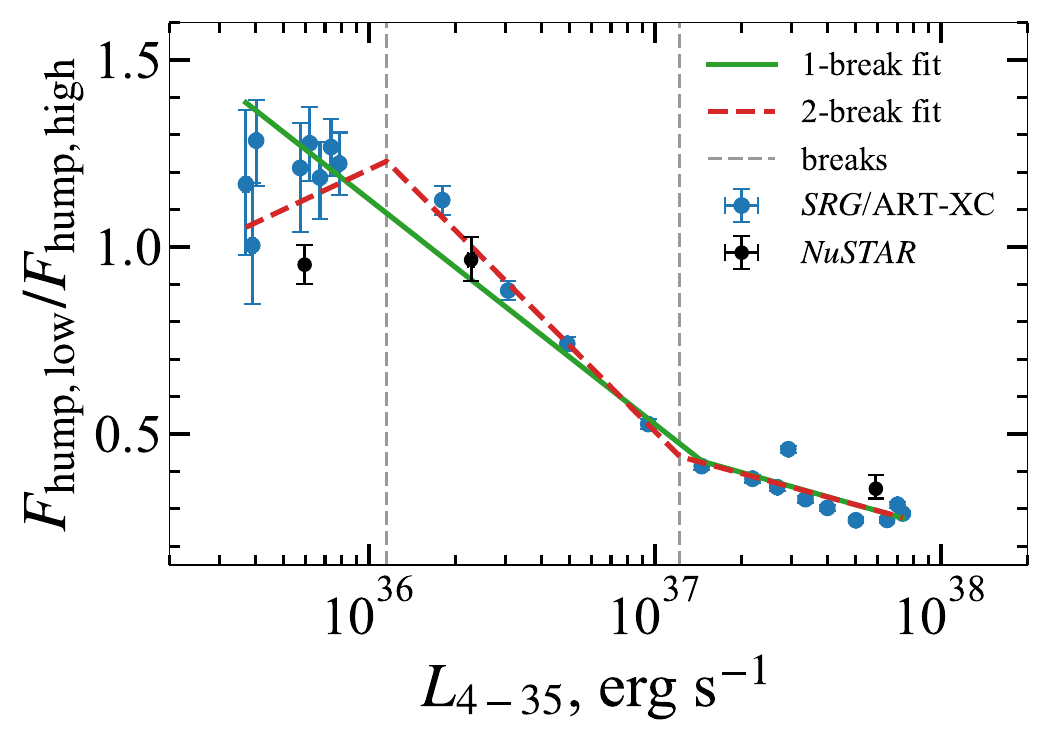}
\caption{Flux ratio of the low- and high-energy continuum humps as a function of 4--35~keV luminosity. 
Blue and black points show the individual \srg/ART-XC and the three \nustar measurements, respectively. 
The green solid curve shows a continuous piecewise log-linear fit with one free break. 
The red dashed curve shows the corresponding fit with two free breaks. 
The vertical dashed lines mark the best-fit break positions of the
two-break model. }
\label{fig:luminosity_vs_ratio}
\end{figure} 

We also examined the model-dependent ratio between the two continuum components, $F_{\rm low}/F_{\rm high}$. For each ART-XC observation, the low- and high-energy \texttt{comptt} fluxes were computed from the best-fit model over 3--79~keV, a band selected to better cover the high-energy hump, which peaks at $\sim$30--40~keV and may extend beyond the nominal ART-XC fitting range. Because ART-XC does not directly constrain the full 3--79~keV interval, these values partly rely on extrapolation and should therefore be treated as model-dependent estimates. We applied the same calculation to the three \nustar\ observations, for which the full 3--79~keV band is directly covered by the data.

As shown in Fig.~\ref{fig:luminosity_vs_ratio}, the three \nustar\ points lie on the same luminosity trend as the ART-XC measurements, indicating that the extrapolated ART-XC ratios are consistent with the \nustar\ estimates. We therefore fitted the luminosity dependence of this ratio for the combined ART-XC and \nustar\ data set with continuous piecewise log-linear functions with one and two free breaks. The second, low-luminosity break was motivated by the
transition interval $L_{4-35}\simeq(0.8$--$1.8)\times10^{36}$~\lum, where the
high-energy hump becomes resolved in the ART-XC spectra. The models were
compared using the Akaike information criterion (AIC; \citealt{Akaike1974}),
which accounts for the two additional free parameters in the two-break model.

The single-break fit places the break at $L_{\rm b}\simeq(1.5\pm0.1)\times10^{37}$~\lum. 
The corresponding slopes are $-0.60\pm0.02$ and
$-0.21\pm0.01$ below and above the break, respectively.
This suggests that the relative contribution of the two continuum components changes rapidly during the decline below $\sim10^{37}$~\lum, while it evolves more slowly in the bright state.

Allowing for an additional low-luminosity break substantially improves the
description of the data. The AIC favors the two-break model, with
$\Delta{\rm AIC}=-43$ relative to the single-break model. The two-break fit gives breaks at
$L_{\rm b,1}=(1.2^{+0.4}_{-0.2})\times10^{36}$~\lum\ and
$L_{\rm b,2}=(1.2\pm0.1)\times10^{37}$~\lum. 
The corresponding slopes are $0.4\pm0.3$, $-0.78\pm0.04$, and $-0.21\pm0.01$ in the three luminosity segments.  
We therefore conclude that the luminosity dependence of $F_{\rm low}/F_{\rm high}$ shows a change in slope in the same luminosity range where the high-energy hump becomes resolved, indicating a change in the relative behavior of the two continuum humps at this luminosity.

\section{Discussion}
\label{sec:discussion}

\subsection{Luminosity-dependent strengthening of the high-energy hump}
\label{sec:HEH_theory}

The broad-band X-ray spectra of accreting XRPs at typical outburst luminosities are generally dominated by a hard continuum, canonically modeled as a power law with an exponential high-energy cutoff. 
This classical spectral shape is thought to result from Comptonization of seed photons in the hot plasma of the accretion column, with contributions from thermal and bulk motions of the infalling material \citep[see][for a recent review]{MushtukovTsygankov2024}. 

At luminosities from $L_{\rm X}\sim10^{34}$~\lum\ \citep{Tsygankov2019b, Salganik2025, Malacaria2026, Zalot2026} to $\sim10^{36}$~\lum\ \citep{Salganik2023}, several long-period XRPs developed a more complex two-component spectral shape, in which a distinct high-energy hump appears alongside the lower-energy continuum component. 
In \src, this transition is directly traced by the dense ART-XC monitoring and the three broad-band \nustar observations: as the source fades, the spectral shape evolves from a cutoff power-law continuum into a pronounced two-hump morphology, with broad maxima near $\sim$10~keV and $\sim$30--40~keV. 
The high-energy hump becomes clearly resolved within $L_{4-35}\simeq(0.8$--$1.8)\times10^{36}$~\lum. 
A similar behavior was observed in the XRP A~0535+262 \citep{Tsygankov2019a}, which has a comparable CRSF energy and therefore a similar magnetic-field strength in the line-forming region.

As the source brightens, the high-energy hump becomes stronger relative to the
low-energy component in the 4--35~keV band
(see Fig.~\ref{fig:luminosity_vs_ratio}). This suggests that the relative
strength of the two spectral components is not fixed, but depends on the
conditions in the accretion flow.

In the low-accretion-rate regime, braking of the plasma in the NS atmosphere produces cyclotron photons and partially thermalizes the kinetic energy of the accretion flow. 
The resulting thermal radiation and cyclotron photons, after multiple scatterings in the atmosphere, form the two spectral components observed at low mass accretion rates \citep[see, e.g.,][]{Mushtukov2021,SokolovaLapa2021}.

In \src, the increasing relative strength of the high-energy hump may indicate that the radiation produced in the NS atmosphere is additionally reprocessed in the accretion flow above the polar cap.
Photons produced close to the stellar surface have to propagate through the infalling plasma above the polar regions. 
Below the cyclotron energy the flow remains optically thin, but near the cyclotron resonance the optical depth becomes large at relatively low mass accretion rates.

Therefore, photons with energies comparable to, or exceeding, the local cyclotron energy can resonantly interact with the infalling plasma \citep{Mushtukov2015b,2026arXiv260202733F,2026arXiv260506162M}. 
In the rest frame of the rapidly moving electrons, photon energies are Doppler-shifted upward, making resonant scattering possible over an extended part of the accretion flow. 
Such scatterings can excite electrons to higher Landau levels, whose subsequent radiative de-excitation produces photons near the fundamental cyclotron energy \citep{1986ApJ...309..362D,2016PhRvD..93j5003M}.

This is the same physical process usually referred to as photon spawning in cyclotron-line formation calculations \citep{1999ApJ...517..334A}.
In this picture, high-energy photons are not simply removed from the spectrum, but are redistributed toward the cyclotron energy. 
The spawned photons may be partly directed toward the NS surface, reprocessed in the atmosphere, and scattered back into the accretion flow, enabling further resonant interactions. 
As the mass accretion rate increases, the resonant optical depth of the accretion channel also increases, making this reprocessing more efficient. 
This provides a natural qualitative mechanism for enhancing the high-energy hump around the cyclotron energy and suppressing the broader high-energy tail above the resonance.

\subsection{Possible transition to the accretion-column regime}
The change in the relative contribution of the two spectral humps may also carry
information about the accretion regime.
In particular, the evolution of the hump-flux ratio shown in Fig.~\ref{fig:luminosity_vs_ratio} may indicate the presence of a characteristic luminosity scale around  $L_{4-35} \sim 1 \times 10^{37}\,{\rm erg\,s^{-1}}$. 
At lower luminosities, the ratio $F_{\rm low}/F_{\rm high}$ changes rapidly with luminosity, whereas above this level the evolution becomes noticeably slower. 
This luminosity range is broadly consistent with the expected onset of radiative braking in the accretion flow above the NS surface \citep{BaskoSunyaev1976}. 
As radiation pressure becomes increasingly important, the bulk velocity of the infalling plasma may decrease before the flow reaches the NS atmosphere. 
In this case, accreting particles would deposit their kinetic energy at lower velocities, potentially reducing the efficiency of electron excitation to higher Landau levels and weakening the luminosity dependence of the high-energy hump formation processes discussed in Sect.~\ref{sec:HEH_theory}.

At higher mass accretion rates, this same transition may eventually lead to the formation of a radiative accretion column. 
In such a regime, part of the radiation produced in the column can be reflected and reprocessed by the NS atmosphere \citep{2013ApJ...777..115P}, which may soften the observed continuum and further reduce the rate at which the spectral hardness increases with luminosity \citep{2015MNRAS.452.1601P}. 
A qualitatively similar effect may also arise if the accretion flow becomes increasingly optically thick in the continuum.

\section{Conclusions}
\label{sec:conclusions}

We presented a dense X-ray monitoring campaign of the 2026 outburst of the
Be/X-ray pulsar \src, using \srg/ART-XC, \nustar, and \hxmt observations.
Our main results can be summarized as follows:

\begin{itemize}

\item The 2026 outburst of \src was followed from the bright state down to the
low-luminosity tail. The \srg/ART-XC monitoring traces a smooth decay from
$L_{4-35}\gtrsim10^{37}$~\lum\ to a plateau at
$L_{4-35}\simeq(3$--$8)\times10^{35}$~\lum.

\item The broad-band \nustar spectra, modeled with a two-component thermal
Comptonization continuum, show a clear luminosity-dependent evolution. At high
luminosity, the combined continuum resembles the standard cutoff power-law spectral
shape commonly observed in accreting X-ray pulsars. At lower luminosities, the high-energy hump becomes distinct, and the spectrum
shows two broad components: a low-energy hump peaking near $\sim$10~keV and a
high-energy hump peaking near $\sim$30--40~keV. 

\item Using the ART-XC monitoring, we constrain the luminosity range over which
the high-energy hump becomes clearly resolved to
$L_{4-35}\simeq(0.8$--$1.8)\times10^{36}$~\lum.

\item At higher luminosities, the flux ratio between the low- and high-energy
humps shows a break at $L_{4-35}\sim1\times10^{37}$~\lum, which may indicate a
characteristic luminosity scale associated with the onset of radiative braking
in the accretion flow.

\item The CRSF at
$E_{\rm cyc}\sim55$~keV is observed in the high-energy hump. Its
centroid energy does not show a significant luminosity dependence over the
observed luminosity range.

\end{itemize}

\begin{acknowledgements}
 
AS acknowledges support from the Jenny and Antti Wihuri Foundation (grant 00240331).
This research was supported by the International Space Science Institute (ISSI) in Bern, through International Team project 25-657 `Polarimetric Insights into Extreme Magnetism’ and the  Research Council of Finland Centre of Excellence in Neutron-Star Physics (grant 374064). 
SST, SVM, IYL, AAL, AYT, and JP acknowledge support from the Ministry of Science and Higher Education of RF grant 075-15-2024-647.
AAM acknowledges support from the UKRI Stephen Hawking fellowship.
HX acknowledges support from the China Scholarship Council (CSC).

This work is partially based on observations with the Mikhail Pavlinsky ART-XC telescope, hard X-ray instrument on board the \srg observatory. The \srg observatory was created by Roskosmos in the interests of the Russian Academy of Sciences represented by its Space Research Institute (IKI) in the framework of the Russian Federal Space Program, with the participation of Germany. The ART-XC team thanks the Roscosmos State Corporation, the Russian Academy of Sciences, and Rosatom State Corporation for supporting the ART-XC telescope, as well as the JSC Lavochkin Association and partners for manufacturing and running the Navigator spacecraft and platform.  

This work also made use of data obtained with the NuSTAR mission, a project led by Caltech, funded by NASA, and managed by JPL, and data from the \hxmt mission, a project funded by the China National Space Administration (CNSA) and the Chinese Academy of Sciences (CAS). This research also has made use of the \nustar Data Analysis Software (NUSTARDAS) jointly developed by the ASI Science Data Centre (ASDC, Italy) and Caltech.

We are grateful to the \hxmt, \srg, and \nustar teams for approving and rapid scheduling of the monitoring campaign.

\end{acknowledgements}

\bibliography{allbib}
\bibliographystyle{aa}
\clearpage
\appendix
\onecolumn
\section{Observation log}
\label{app:obslog}

Table~\ref{table:obslist} summarizes the \hxmt\ and \srg/ART-XC observations of \src\ used in this work and lists the corresponding spectral-fit results. 
For each observation we list the observation identifier, start and end times in MJD, the effective exposure after standard screening, and the main parameters derived from the spectral fits.
The three \nustar\ observations are summarized separately in Table~\ref{tab:specs}.

\begin{table}[h]
\setlength{\tabcolsep}{3pt}
\centering
\caption{Observation log and spectral-fit parameters for the \srg/ART-XC and \hxmt\ observations.}
\label{table:obslist}
\begin{tabular}{lccccccccccc}
\hline\hline
ObsID & $T_{\rm start}$ & $T_{\rm stop}$ & Exp. 
 & $kT_{0}$ & $kT_{\rm e,low}$ & $kT_{\rm e,high}$ 
& $\tau_{\rm low}$ & $E_{\rm cyc}$ & $L_{4-35}$ & $L_{3-79}$ & $\chi^2$/d.o.f. \\
 & MJD & MJD & ks 
& keV & keV & keV & & keV 
& \multicolumn{2}{c}{$10^{35}$~\lum} & \\
\hline
\multicolumn{11}{l}{\hxmt} \\
P0804763002 & 61061.4 & 61061.8 & 34.8
& 1.24 & $4.6\pm0.4$ & $7.6\pm0.4$ & $5.8\pm0.1$ & $55\pm1$ & $340\pm2$  & $402\pm2$ & 1232/1191 \\ 
P0804763003 & 61062.4 & 61062.8 & 34.1
& 1.24 & $2.8\pm0.1$ & $6.8\pm0.1$ & $9.2\pm0.3$ & $53.9\pm0.6$ & $367\pm1$ & $437\pm1$ & 1165/1191 \\
P0804763004 & 61063.8 & 61064.0 & 17.7 
& 1.25 & $3.2\pm0.1$ & $6.9\pm0.1$ & $7.2^{+0.3}_{-0.2}$ & $51.4\pm0.5$ & $555\pm4$ & $655\pm4$ & 1139/1188 \\
P0804763005 & 61064.4 & 61064.8 & 34.8
& 1.25 & $3.1\pm0.1$ & $6.8\pm0.1$ & $7.5\pm0.2$ & $53.9\pm0.4$ & $603\pm3$ & $712\pm4$ & 1114/1191 \\
P0804763006 & 61065.4 & 61066.0 & 51.6
& 1.25 & $3.2\pm0.1$ & $6.8\pm0.1$ & $7.4\pm0.2$ & $54.3\pm0.3$ & $708\pm3$ & $830\pm3$ & 1149/1191 \\

P0804763007 & 61066.6 & 61067.1 & 40.1
& 1.25 & $2.8\pm0.1$ & $6.8\pm0.1$ & $9.3\pm0.3$ & $53.8\pm0.3$ & $663\pm4$ & $778\pm5$ & 1137/1191 \\
P0804763008 & 61067.5 & 61067.9 & 35.1
& 1.25 & $3.3\pm0.1$ & $6.8\pm0.1$ & $6.9\pm0.2$ & $54.6\pm0.3$ & $766\pm4$ & $895\pm5$ & 1077/1191 \\
\hline

\multicolumn{11}{l}{\srg/ART-XC} \\
12610180001 & 61059.5 & 61062.3 & 151.6
& 1.23 & 3.58 & $6.8\pm0.1$ & $6.5\pm0.2$ &  -- & $291.7\pm0.5$ & $360\pm2$ & 44/124 \\
12610180002 & 61064.5 & 61065.4 & 79.6
& 1.25 & 3.60 & $6.7\pm0.1$ & $6.6\pm0.2$ & --  &  $702\pm1$  & $876\pm4$ & 41/124 \\

12610180003 & 61066.5 & 61068.6 & 173.7
& 1.25 & 3.60 & $6.7\pm0.1$ & $6.6\pm0.2$ & -- & $733\pm1$ & $916\pm4$ & 42/124 \\
12610180004 & 61069.6 & 61070.6 & 89.3
& 1.25 & 3.60 & $6.6\pm0.1$ & $6.3\pm0.2$ & -- & $646\pm1$ & $802\pm4$  & 42/124 \\
12610180005 & 61072.8 & 61073.2 & 41.3
& 1.25 & 3.60 & $6.6\pm0.1$ & $6.2\pm0.2$ & -- & $502\pm1$ & $624\pm4$ & 50/124 \\ 
12610180006 & 61074.5 & 61075.0 & 41.9 & 1.24 & 3.59 & $6.7\pm0.1$ & $6.2\pm0.2$ & -- & $399.5\pm0.9$ & $499\pm3$ & 51/124 \\

12610180007 & 61075.9 & 61076.4 & 44.1
& 1.24 & 3.58 & $6.6\pm0.1$ & $6.1\pm0.2$ & -- & $334.9\pm0.8$ & $415\pm3$   & 46/124 \\
12610180008 & 61077.7 & 61078.2 & 44.2
& 1.23 & 3.58 & $6.7\pm0.1$ & $6.1\pm0.2$ & -- & $266.8\pm0.7$ & $332\pm3$ & 55/124 \\
12610180009 & 61079.1 & 61079.6 & 44.5
& 1.23 & 3.57 & $6.7\pm0.1$ & $5.9\pm0.2$ & -- & $218.6\pm0.6$ & $270\pm2$ & 59/124 \\
12610180010 & 61081.5 & 61082.0 & 45.5
& 1.22 & 3.56 & $6.4\pm0.1$ & $5.3\pm0.2$ & --  & $145.2\pm0.4$ &  $176\pm1$ & 88/124 \\

12610180011 & 61082.9 & 61083.9 & 83.4
& 1.21 & 3.54 & $6.5\pm0.1$ & $5.3\pm0.1$ & -- & $94.5\pm0.3$ & $114\pm1$ & 57/124 \\

12610180012 & 61086.1 & 61087.0 & 83.2
 & 1.20 & 3.52 & $6.7\pm0.1$ & $5.5\pm0.1$ & -- & $49.2\pm0.2$ & $59.8\pm0.6$ & 99/124 \\

12610180013 & 61089.2 & 61090.2 & 84.4
 & 1.19 & 3.51 & $6.8\pm0.2$ & $5.3\pm0.1$ & -- & $30.6\pm0.1$ &  $37.1\pm0.5$ & 105/124 \\
12610180014 & 61094.1 & 61095.1 & 87.2
& 1.15 & 3.45 & $6.9\pm0.2$ & $5.3\pm0.1$ & -- & $18.0\pm0.1$ &  $21.8\pm0.4$ &
104/124 \\
12610180015 & 61098.3 & 61099.3 & 82.7
& 1.02 & 3.26 & $7.9\pm0.5$ & $5.5\pm0.1$ & -- & $7.9\pm0.1$ &  $10.1\pm0.4$& 136/124 \\
12610180016 & 61102.1 & 61103.0 & 81.0
& 0.98 & 3.20 & $10^{+2}_{-1}$ & $5.8\pm0.1$ & -- & $3.7\pm0.1$ & $5.3^{+0.3}_{-0.2}$ & 131/124 \\
12610180017 & 61105.8 & 61106.8 & 84.0
& 0.98 & 3.20 & $10^{+2}_{-1}$ & $5.7\pm0.1$ & -- & $3.9\pm0.1$ & $5.8^{+0.8}_{-0.6}$ & 100/124 \\
12610180018 & 61109.6 & 61111.6 & 171.2
& 0.98 & 3.20 & $8.6^{+0.7}_{-0.5}$ & $5.6\pm0.1$ & -- &  $4.0\pm0.1$ &  $5.4^{+0.3}_{-0.2}$& 110/124 \\
12610180019 & 61115.0 & 61116.6 & 138.4
& 1.01 & 3.25 & $8.0\pm0.4$ & $5.6\pm0.1$ & -- & $7.4\pm0.1$ &  $9.5\pm0.3$ & 122/124 \\
12610180020 & 61120.2 & 61121.3 & 92.3
& 1.00 & 3.23 & $8.9^{+0.8}_{-0.5}$ & $5.8\pm0.1$ & -- & $6.7\pm0.1$ & $9.1^{+0.6}_{-0.4}$ & 111/124 \\
12610180021 & 61125.5 & 61126.6 & 112.2
& 0.99 & 3.21 & $8.2\pm0.6$ & $5.7\pm0.1$ & -- & $6.2\pm0.1$ &  $8.1^{+0.4}_{-0.3}$ & 135/124 \\
12610180022 & 61131.0 & 61131.8 & 65.6
& 0.98 & 3.20 & $8.7^{+0.9}_{-0.8}$ & $5.8\pm0.1$ & -- & $5.8\pm0.1$ & $7.8^{+0.9}_{-0.4}$ & 138/124 \\

\hline
\end{tabular}
\tablefoot{
The continuum was described with the two-component Comptonization model discussed in Sect.~\ref{sec:spectrum}. 
For all fits, the optical depth of the high-energy Comptonization component was fixed at $\tau_{\rm high}=200$, since it was not independently constrained.
For the \hxmt\ fits, the CRSF width was fixed at $\sigma_{\rm cyc}=10.7$~keV.
For the ART-XC monitoring spectra, we used the simplified model; $kT_{0}$ and $kT_{\rm e,low}$ were fixed to
luminosity-dependent values interpolated from the three \nustar measurements, see Sect.~\ref{sec:artxc_lightcurve}.
For the \hxmt\ fits, only $kT_{0}$ was fixed to interpolated values.
}

\end{table}

\end{document}